# Power Delivery of the Future

Mario Rabinowitz
Armor Research,  715 Lakemead Way, Redwood City, CA 94062-3922
Mario715@earthlink.net

**Abstract**

This paper is written to provide an insight into the physics and engineering that go into power delivery of the future.  Topics covered are Fault Current Limiters (FCL) including Superconducting FCL and Emission Limited FCL; Lightning and Restoration Preparedness; Compressed-Gas-Insulated Delivery; Evaporative Cooling  Delivery; Advanced Delivery Technologies Requiring Big Breakthroughs such as Conducting Polymers, Electron-Beam Delivery, Microwave Delivery, and Laser-Beam Delivery.

**Fault Current Limiters (FCL)**

*General*

As power systems have grown, the need has developed for a fault current limiter (FCL) to keep potentially excessive fault currents within the ratings of existing equipment such as circuit breakers and transformers.  Small FCL (some using little explosive cartridges for fast response to a fault condition) have been developed with reset times of the order of a minute.

In 1974,  a consensus of electric utilities listed the need for a FCL as a top priority R & D item.  Almost all of the devices that were considered 20 years ago were either technically or economically unsuccessful.  Such devices fell into two broad categories.  In the first category, a tuned LC circuit in which the inductive reactance, is used in series in a power line to give a low impedance at the power frequency.  When a fault (i.e. a short circuit) occurs, a switch shorts out the capacitor (C), and the inductive reactance limits the current.  The disadvantages of this approach include large size, big capital cost, and high operating costs.

In the second category, an impedance in parallel with a normally closed bypass switch is placed in series in the power line.  When a fault is sensed, the bypass switch is opened and the current is transferred to the current limiting impedance.  The approaches tried have included unstable vacuum arcs controlled by a magnetic field or other high arcing voltage circuit breakers in parallel with resistors; switches in parallel with fuses in parallel with resistors; and driving superconductors into their resistive normal state.  Some of the many disadvantages are related to the difficulty in switching, and slow reaction time because of the time required for sensing and switching operations.

Some solid state diodes have very low conductivity with rapid current saturation at low voltage when operated in the reverse bias mode. Putting two such diodes in opposing back-to-back direction, in series, gives practical current limitation for low voltages. However, they are not yet practical for power applications for tens of kilovolts.

*Superconducting FCL*

Exciting new developments in fault current limiters using high temperature superconductors (HTS) show promise for use in future power delivery. In one design, a superconducting shield surrounds and shields an iron core from the magnetic field of a normal coil so that the core looks like an air core. The normal coil is in series with the power line. The superconducting shield is driven normal when there is a fault, greatly increasing the coil's inductive reactance by the permeability of the iron core.

In another approach referred to as a differential-mode fault limiter, the FCL consists of a normal primary high voltage coil coupled to an HTS secondary coil. The fault current only passes through the normal coil which is in series with the power line. The primary coil is linked to the superconducting coil electromagnetically. The HTS stays in the superconducting state during a fault. When a fault current appears, the inductive coupling between the two coils increases the inductance of the primary by a factor of 60. Thus superconducting fault current limiters that not only protect the system but also ease the burden on circuit breakers are a definite possibility.

*Emission Limited FCL*

A novel FCL design for which two patents have been granted, relates to emission limited fault current limiters [1, 2]. The FCL controls the current thermionically, by cold cathode emission, or with plasma devices. The common feature is that the current is limited by controlling a series emission process. The apparatus can operate independently of a fault current sensor, and limit a fault current automatically by the physics of its operation.

The apparatus is in series in the circuit in conjunction with a cathode from which electrons emanate, and which at a predetermined current level becomes emission limited. This yields a saturation current between the cathode and anode independent of the voltage across the device, thus limiting the current. In a thermionic emitter, it is the temperature of the cathode which determines the saturation or emission limit of the current.

The physics of operation is straightforward. Even though as the voltage increases, the electrons go at a higher velocity as they traverse the gap, the current does not increase. This is because as electron speed increases, the volume number density of the electrons decreases inversely proportional to the speed, thus keeping the current constant. Such devices have potential in the distribution system. Cesium plasma triodes should also be considered in this connection as FCL or power switches. The electron current that passes into the grid-anode region may be emission limited from the cathode, space charge limited from the cathode, or grid limited from the cathode.

**Lightning and Restoration Preparedness**

The National Lightning Detection Network (NLDN) developed and now sponsored by EPRI can aid in the scheduling of restoration preparedness of a system before a major storm by providing lead time for allocation of resources such as repair crew scheduling and dispatch. Potential storm severity can be assessed leading to a more efficient use of manpower. During the storm it can determine areas of potential heavy damage. After the storm it can aid in dispatching restoration personnel.

In lightning-prone areas, lightning is the largest cause of power failures accounting for about half of the outages, and costing utilities about $100 million to $200 million per year just to replace damaged equipment. Although individual lightning strikes usually do not cause severe damage, their combined annual effect is disastrous. There are roughly 15 million strikes per year in the U.S., killing several hundred people. Worldwide more people are killed or injured each year by lightning than by tornadoes, floods, and hurricanes combined. Lightning has also caused forest fires, extensive property damage, serious nuclear power plant malfunction, and disruption of the navigational devices on commercial airliners and spacae rockets. About 65 % of all kinds of equipment damage during thunderstorms is directly caused by lightning.

Lightning is made up of one or more separate strokes, which are intermittent partial discharges. Peak currents greater than 500,000 A have been directly measured. Currents measured at the ground can rise to 20,000 A in less than a microsecond, with a maximum time rate of change of current ~ $10^{11}$A/sec. Lightning creates electric fields in the discharge region greater than $10^6$ V/m, with power levels ~ $10^{12}$ W , and energy dissipation between $10^9$ and $10^{10}$ Joules.  The potential of charged clouds that produce lightning can be as high as a billion volts.  The current channel has a sharp temperature rise due to the rapid energy input.  This causes the channel to expand with supersonic speed in times less than 10 microseconds, producing the roughly cylindrical shock wave which produces thunder.  The channel temperature, as measured spectroscopically by Doppler broadening is about 30,000 K.  By all standards, lightning is quite a violent phenomenon with which the utility sytem has to cope regularly.

Lightning transients can propagate along transmission lines at almost the speed of light, with circuit-limited rise times of about a microsecond to peak voltages as high as millions of volts, with a maximum rate of rise of ~ $10^{12}$ Volts/sec.  For example, a lightning strike on the 110-kV line of the Arkansas Power & Light Company reached a peak voltage of 5 million volts within 2 microseconds as measured by their oscillograph.  Such direct lightning strikes can produce the most severe effects.  The high current density ~ $10^3$ A/cm$^2$ in a stroke delivers a high power density to the strike point resulting in demolished structures such as exploded timber, molten metal, and charred insulation.

Electromagnetic field data generated by lightning strikes is detected by 105 NLDN sensors.  This data is relayed to the Network Control Center (NCC) in Tucson, which computes and transmits the magnitude, time, and position of the strikes. Operators of this network can pinpoint the location and magnitude of individual strikes by measuring the direction and time of the electromagnetic pulses given off by lightning.  Subscribers receive this data within seconds of the strike time, and the lightning activity is shown on regional maps.  Time is measured with great precision to within microseconds.  This can translate to potential location precision within a fraction of a mile given by longitude and latitude.  Magnitude is given as a rough estimate of the peak stroke current in kA which relates to the protection needed by a given line.  The distribution of peak currents and density of strikes are important in power system design.

Timing, directional, and lightning strength sensors provide input to the NLDN. Signal strength, azimuth, polarity, and time data are obtained for the first stroke, combined with the number of additional strokes.  Information from 105 sensors is time-correlated.  When enough sensors detect a given stroke, a location can be determined by the X.25 processor.  For real-time users, stroke information is pictorialized into flash information superimposed on a map on a computer screen.  This information can also be downloaded as transcribed data.

Thus lightning fault analysis and location system software provide utilities with a workstation tool enabling them to determine the simultaneous occurrence of a lightning event and either a fault, momentary, or other power disturbance. It is able to locate the event to within 500 m. In case of a fault, use of a modern fault finder could increase the resolution to within 5 m. This will facilitate real-time fault location. This system can also be used for retrospective analysis of line performance in lightning storms, and determine which lines or line sections need upgrade. This approach will minimize utility upgrade investment costs. The combination of the Network for restoration preparedness and the workstation for disturbance location and time performance analysis should be a vital part of future power system strategy and tactics.

**Compressed-Gas-Insulated Delivery**

Although no new major developments are needed for compressed-gas-insulated (CGI) transmission lines, and they were first introduced into service in 1971, CGI lines remain a system of the future. To date, relatively few CGI systems have been installed, and these have been mainly in substation getaways. The main insulation is sulfur hexafluoride ($SF_6$) gas, at about 4 atmospheres (400 kPa) pressure. A hollow Al inner conductor is periodically supported by solid insulators from a concentric Al sheath.

Cleanliness and absence of contaminants are very important in maintaining high dielectric strength in CGI systems. In addition to highly polishing the conductor, and assembly under very clean conditions, electrostatic particle traps are provided near the solid insulators to catch and hold any particulate matter which may be left over or form in the lines. Both the degree of installation cleanliness and efficacy of the particle traps have reached a level that results in high integrity of the insulation.

The majority of existing CGI lines have been under a third of a mile in length, at voltages ranging from 145 kV to 550 kV. An 800 kV prototype system has been developed, and work was done on a 1200 kV CGI system. In the U.S., Westinghouse pursued a rigid system, and ITE-Gould offered either a rigid or flexible system using corrugated Al. CGI systems were more actively pursued in Japan.

One of the arcing by-products of $SF_6$ is $S_2F_{10}$ ( equivalently $SF_5$) which is one of the most poisonous gases known. It seems not to be a concern of the EPA, possibly because it is very reactive and the small amounts formed quickly combine with other compounds in the environment to become non-hazardous. The EPA is looking into the question of banning the use of $SF_6$ as a possible contributor to ozone depletion, With respect to the greenhouse effect, the mass ratio of $SF_6$ production to that of $CO_2$ from coal is ~ 0.1 %. If such a ban goes into effect, one may rule out not only CGI delivery, but it will also impact on $SF_6$ substations and $SF_6$ circuit breakers. Fluorocarbons, previously used as coolants in refrigerators, were banned in the U.S. as contributors to depletion of the atmosphere's ozone layer. It is important to anticipate EPA actions in this domain.

**Evaporative Cooling Delivery**

Within the context of conventional technology, evaporative cooling delivery offers some attractive possibilities. The main objective is to increase the cooling rate at the inner conductor, where it is most needed. Another objective is to cool the conductor by a method that requires the least flow rate, viscous loss, and pressure drop of the coolant. This may be achieved by the use of liquids having a high latent heat of evaporation. This would allow large cooling rates with relatively small mass flow of coolant, and hence relatively small hydrodynamic losses.

The drawback of evaporative cooled lines is that they do not increase power delivery capability by a reduction of power losses, but rather increase it by more efficiently removing the power losses from the conductor to keep it from overheating. Thus increased power density is attainable, but at the price of increased power losses.

**Advanced Delivery Technologies Requiring Big Breakthroughs**
*Conducting Polymers (CP)*

A metal easily conducts electricity because a significant fraction of its electrons move freely from atom to atom with long mean free paths between collisions. A number of organic polymers, in which the outer orbital shells overlap extensively, can be made to mimic a metal. In essence they are somewhat like a metal. These polymers are generally quasi-one-dimensional in that they are composed of fine molecular filaments or fibrils, which thread the solid together much like a woven fabric or rope. Along the axes of the threads, the conductivity is high. It is correspondingly low in any non-axial direction.

Conducting polymers (CP) are basically large energy gap, quasi-one-dimensional semiconductors. The quasi-one-dimensional or chain nature results in many unusual properties, such as spinless current carriers. These polymers can be made highly conducting, even metallic by the addition of suitable (usually nonmetallic) impurities. With proper doping, these polymer crystals act like synthetic metals. Because of their reasonably high conductivities, low weight densities, and high tensile strength, the specific (normalized) conductivity of these materials exceeds that of copper and aluminum. Among such polymers are polyacetylene, polyparaphenylene sulfide, anthrone polymers, and polypyrrole. Many questions regarding their chemical stability, electrical, thermal, mechanical, flammability, toxicity and other properties must be answered before such esoteric materials can be used for the transmission of electricity. Foremost among these questions is the one pertaining to chemical stability. Despite being one of the first CPs, having a very high doped electrical conductivity, and very nonlinear optical properties, polyacetylene in both the cis and trans phases remains largely a laboratory curiosity because it is unstable in air. The toxic dopants, such as iodide compounds and arsenic pentafluoride, tend to leach out of the polymers. Without the dopants the polymers lose their high conductivity.

Recent attention has been focused on the optical properties of conducting polymers because some of these polymers, notably polyphenylene vinylene (PPV) have shown promise as light-emitting diodes and, extrapolating from there, flat panel displays for computers, etc. Polythiophene is a more stable molecule that finds uses in solar cells and thin film transistors. PITN is a polythiophene derivative that is quite transparent in its highly conductive state suggesting use as a transparent electrode, and/or in solar cells. Polyaniline's absorption of electromagnetic radiation can be controlled by changing its doping level and local structure. The DAD molecule can act as an optical molecular switch that is light intensity dependent with switching speeds $\sim 10^{-12}$ sec. Polyquinoline molecules have very high thermal stability and may be used in xerography as photoconductors. New developments in the CP arena may well enhance the outlook for their use as a conductor in cables. Even if they are not used as the main conductor, they may have application as the low conductivity cable sheath which is presently made of carbon particle impregnated polymer such as cross-linked polyethylene.

**Electron-Beam Delivery (EBD)**

Two patents filed by Nicholas Christofilos [3,4] in 1950 and 1956 form the basis of electron-beam delivery (EBD). This is a method of electric power delivery in which electricity is transmitted in kinetic form by a magnetically focused electron beam in an evacuated pipe. *While the beam is in transit, this is the closest thing to room temperature superconductivity known to man.* However, the return path is in the pipe wall which is just an ordinary conductor, so only half the line has the advantage of having virtually no resistance.

The advantage of electron-beam delivery is not in its efficiency, as the total power losses may be as large or larger than in a conventional line. EBD may have a capital cost advantage at very high power levels ≥ 10,000 MVA since high-voltage insulation is not required over most of the line. Basically magnetic shielding replaces electrical shielding (insulation), as the beam is transmitted at ground potential.

Electrons are injected from a cathode which is at a large negative potential with respect to a virtual anode at ground potential. They are focused through a hole in the center of the virtual anode and drift at a high velocity inside an evacuated spiral quadrupole focused delivery tube. The quadrupole focusing facilitates bending the electron beam around corners. The kinetic energy of the electrons is recovered at the collection end where they enter a retarding electric field as they slow down and finally come to rest. Thus the electron kinetic energy is converted to potential energy when the electrons are collected on the true anode, and is available to do work on the load.

One problem with EBT is that even a narrow energy spread in the electron beam can translate into large losses at the collector. Maintenance of the needed high vacuum pressure of less than $10^{-7}$ Torr will not be easy. Another problem is that if the electron beam is diverted and strikes the vessel wall, it will burn a hole through not only the tube but any obstacles in its way.

On the other hand, focusing of the electrons is not as difficult as it may seem. In a symmetrical configuration, the electrostatic force of repulsion of the electrons is largely canceled by their self-magnetic field as the velocity of the electron beam approaches the speed of light. In electron-positron storage rings which are similar to electron-beam delivery, the circulating electron bunches have an average power of 200 MW and travel more than a billion miles in just 1 1/2 hours of storage.

*Microwave Delivery*

In 1897 Nikola Tesla applied for (and was granted) a patent in which he claimed that he had invented, and experimentally demonstrated, an advanced form of wireless electric power delivery through the earth and the atmosphere for collection at a distant point. To this day no one has been able to demonstrate this invention. Although the microwave delivery of high power through air sounds a little like Tesla's old dream, the similarity ends with the absence of wires or waveguides.

The efficiency of generating, transmitting, and receiving microwave power appears to be significantly less than by conventional means. At frequencies above 10 MHz, there is a problem of significant beam power absorption by water vapor and oxygen. The sensitivity to water vapor makes microwave delivery vulnerable to rain, snow, and ice. Hence the power capacity and power losses would be quite weather-dependent. This problem could be circumvented by transmitting microwave power inside large cylinders. However, this would greatly add to a highly expensive system which is already burdened by costly and inefficient equipment for conversion from 60 Hz to microwave frequencies and back again to 60 Hz.

Much interest was generated in the last decade in the free-space microwave delivery of power to Earth from an orbiting solar satellite. On top of the power losses and high capital and operating costs, such a free-space system has the grave disadvantage that it may be a danger to humans and animals that get in or near the beam.

*Laser-Beam Delivery*

Someday in the far future, we may have laser-beam power delivery. It would have a clear advantage over ordinary power delivery in that no electrical insulation would be required. It would have a clear advantage over electron-beam delivery in that no magnetic insulation would be required. With two such outstanding advantages, we may well ask: Where's the rub?

The major disadvantage of laser-beam power delivery is in the conversion from electric power to light (electromagnetic radiation) and back again. This process is presently so inefficient that less than $10^{-4}$ of the energy would be available to the load. Even if the light itself were delivered directly to the recipient for illumination purposes, the efficiency is presently too low to be practical. However, solid state devices for direct conversion may someday change these prospects. We may then consider high power laser beams in evacuated vessels for transmission, and fiber optic bundles for

-10-distribution.  The next level of concern will be how to avoid burning holes through the containing vessels.

distribution.  The next level of concern will be how to avoid burning holes through the containing vessels.

**References**

[1] E.J. Britt, G.O. Fitzpatrick, L.K. Hansen, and M. Rabinowitz, U.S. Pat. 4,396,865 (1983)

[2]  M. Rabinowitz and W. H. Esselman, U.S. Pat. 4,594,630 (1986).

[3]  N.C. Christofilos, U.S. Patent 2,736,799 (1956)

[4]  N.C. Christofilos, U.S. Patent 2,953,750 (1960).